\documentclass[oldversion]{aa}
\usepackage[dvips]{graphicx}

\usepackage{subfigure}
\usepackage{rotating}
\usepackage{longtable}
\usepackage{amssymb,amsmath}
\newcommand{\be}{\begin{equation}}
\newcommand{\ee}{\end{equation}}
\usepackage{latexsym}
\newcommand{\simless}{\mathbin{\lower 3pt\hbox
      {$\rlap{\raise 5pt\hbox{$\char'074$}}\mathchar"7218$}}} 
\newcommand{\simgreat}{\mathbin{\lower 3pt\hbox
     {$\rlap{\raise 5pt\hbox{$\char'076$}}\mathchar"7218$}}} 

\newcommand{\um}{$\mu$m}
\newcommand{\Msun}{M$_\odot$}
\newcommand{\Lsun}{L$_\odot$}

\newcommand{\Brg}{Br$\gamma$}
\newcommand{\Teff}{T$_{eff}$}
\newcommand{\Lstar}{L$_\ast$}

\newcommand{\Mstar}{M$_\ast$}
\newcommand{\Macc}{$\dot M_{acc}$}

\begin{document}

\title{Gas and dust in the inner disk of the Herbig Ae star MWC~758
\thanks{
Based on observations collected at the European Southern Observatory, Chile.  
Program 078.C-0243.}
}

\author{
A. Isella\inst{1,2}
E. Tatulli\inst{1,3}
A. Natta\inst{1},
\and
L. Testi\inst{1,4}
}

\institute{
    Osservatorio Astrofisico di Arcetri, INAF, Largo E.Fermi 5, I-50125 Firenze, Italy 
\and
    California Institute of Technology, MC 105-24, 1200 East California Blvd., Pasadena CA 91125, USA
\and
    Laboratoire d'Astrophysique de Grenoble, UMR 5571 Universit\'e Joseph Fourier/CNRS BP 53, F-38041, Grenoble, Cedex 9, France
\and 
    ESO, Karl-Schwarschild Strasse 2, D-85748 Garching bei M\"unchen, Germany
}

\offprints{isella@astro.caltech.edu}
\date{Received ...; accepted ...}

\authorrunning{Isella et al.}
\titlerunning{Inner disk in MWC~758}

\abstract
{In this Letter we investigate the origin of the near-infrared emission of the Herbig Ae 
star MWC~758 on sub-astronomical unit (AU) scales using spectrally dispersed low resolution 
(R=35) AMBER/VLTI interferometric observations both in the H ($1.7 \mu$m) and K ($2.2 \mu$m) bands.
We find that the K band visibilities and closure phases are consistent with the presence of a dusty
disk inner rim located at the dust evaporation distance (0.4~AU) while the bulk of the H band emission 
arises within 0.1~AU from the central star. Comparing the observational results 
with theoretical model predictions, we suggest that the H band emission is dominated by an hot gaseous 
accretion disk.}


\maketitle

\section {Introduction}

The development of long baseline near-infrared (NIR) interferometry allows us 
to spatially resolve the emission of the innermost regions of disks around 
young pre-main sequence stars ($\simless$1 AU in the nearest star forming regions). 
These regions are important for understanding the star-disk interaction, 
the gas accretion process onto the star, and the launching of jets and winds. 
Moreover, planet formation is likely to be favored in the high density 
regions of the inner disk.


The existing NIR interferometric observations of low mass T Tauri and intermediate mass 
Herbig Ae disks (HAe hereafter; Millan-Gabet et al. 2007, Monnier et al. 2006, Akeson et al. 2005, 
Eisner et al. 2004, 2007), show that emission in
the K band (2.2 \um) is dominated by dust at the sublimation temperature, located in a puffed-up rim 
(Dullemond et al.~2001, Isella and Natta~2005). For the more luminous HBe stars, dust emission alone can not explain 
the observed visibilities, and it has been suggested that the K band excess flux is dominated by the emission of a dense, 
optically thick gaseous disk which extends well inside the dust sublimation radius (Malbet et al. 2007).
Hints of the presence of an hot gaseous disk are also provided by recent spectrally dispersed low resolution 
observations of a sample of HAe stars where dust emission alone cannot reproduce the spectral variation
of the visibility across the K band (Eisner et al.~2007).
Most gas tracers, such as the hydrogen recombination lines, are very likely emitted by matter which is either accreting 
or ejected from the system (Malbet et al.~2007; Tatulli et al.~2007; Kraus et al.~2008; Eisner et al.~2007b), rather than from the gaseous disk 
itself. In only one case (51 Oph; Tatulli et al.~2008) Amber/VLTI medium
spectral resolution observations of the CO v=2-0 lines at 2.3 $\mu$m confirm their origin in the
inner disk, inside the dust sublimation radius. 

In this Letter we present spectro-interferometric observations
of the HAe star MWC~758 (ST A5, D=200 pc, \Lstar=22 \Lsun, \Mstar= 2 \Msun)  across the H and the K band 
obtained with AMBER/VLTI. We show that the H and K band excess emission arises from two different disk regions,
the first of which is much more compact and hotter than the second. We identify the dominant source of the  2 $\mu$m 
emission as the dust rim at the sublimation radius, and the dominant source of the 1.7 $\mu$m emission as the gaseous
disk inside this radius.

The structure of this letter is as follows. Sec.~2 describes the observations and the data reduction, 
Sec.~3 describes the results, which are discussed in Sec.~4. 
Sec.~5 summarizes briefly our conclusions.

\section {Observations and data reduction}
MWC758 ($K=5.8$ mag, $H=6.5$ mag) was observed with AMBER on December 30, 
2006, and on January 8, 2007, with three 8~m Unit Telescopes (UTs) of the VLTI, using 
the UT1-UT3-UT4 triplet. On both dates, the source was observed close to  transit; 
the hour angles (0.25 and 0.54 hours, respectively) of the two sets of observations were separated by 20 min.
At the time of the observations, the parameters of the three projected baselines were: B=81m, PA=45$^{\circ}$ (UT1-UT3), 
B=59m, PA=102$^{\circ}$ (UT3-UT4), and B=123m, PA=69$^{\circ}$ (UT1-UT4), 
on Dec.30, 2006,  and  B=84m, PA=45$^{\circ}$ (UT1-UT3), B=58m, PA=101$^{\circ}$ (UT3-UT4), 
and B=125m, PA=68$^{\circ}$ (UT1-UT4), where the  position angle PA is measured east of north. Observations were
performed in low spectral resolution mode (R=35), in both the H and K bands. \\
Raw visibilities and closure phases as a function of wavelength were
computed following the standard data reduction algorithms described in
\cite{tatulli_1}, for each spectral band separately.
The raw visibilities were corrected for atmospheric and
instrumental effects by dividing them by the raw visibilities
measured on a reference source. As a calibrator, we chose HD34053
($K=5.9$, $H=5.9$) which has similar magnitudes and, according to the CADAR catalog (Pasinetti Fracassini 
et al. 2001), has a diameter of  $0.2$~mas  -- that is, unresolved by
the interferometer. The closure phase is theoretically independent on the Earth's atmosphere
(see, e.g., Monnier et al. 2007 for the definition of closure phase) and does not 
need to be calibrated. However, AMBER introduces an offset in 
the closure phase which is corrected by subtracting the calibrator closure phase from the
MWC758 observed values. 

\section {Results}

\begin{figure*}[!t]
	\centering
	\subfigure[]
		{\includegraphics[angle=0, width=9cm]{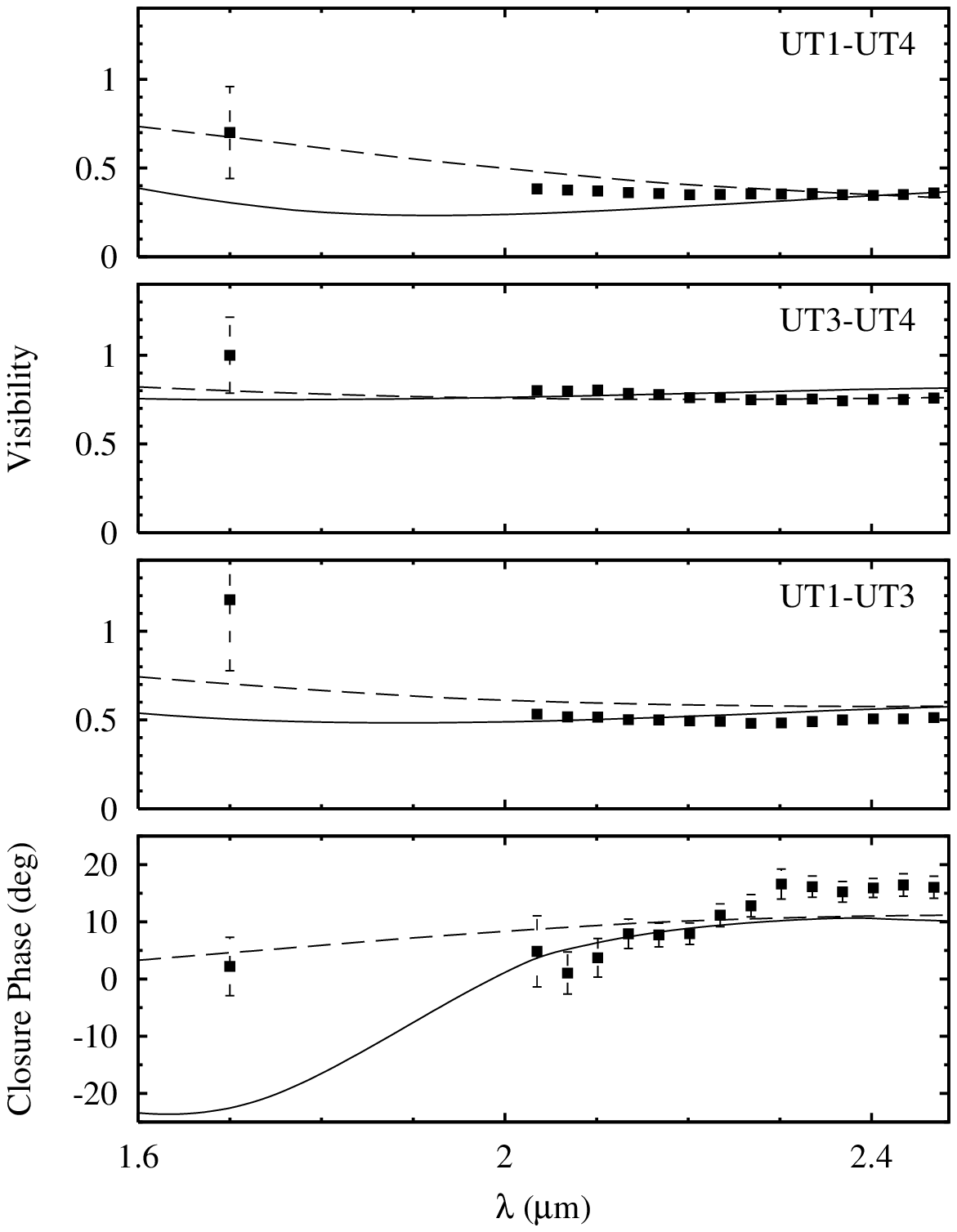}}
	\subfigure[]
		{\includegraphics[angle=0, width=9cm]{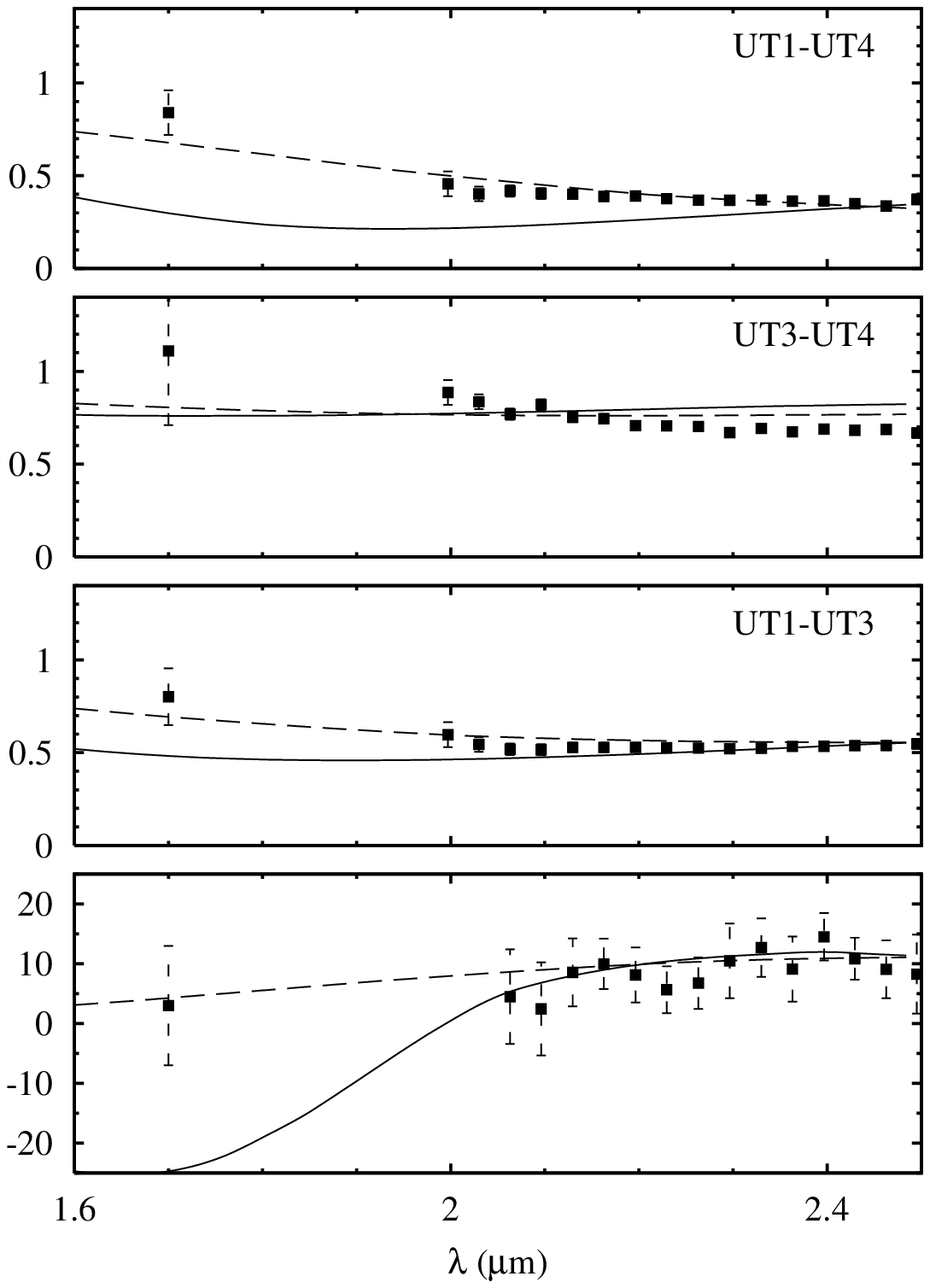}}
	\caption{Spectrally dispersed visibility and closure phase measured across the H (1.65~$\mu$m) and K (2.2~$\mu$m) bands.
	Panels (a) and (b) show the data for two different hour angles, HA=0.25 h and HA=0.54 h respectively. The three upper panels show
	calibrated visibilities for the three different baselines as labeled in figure and described in Sec.~2; corresponding closure phases
	are plotted in the bottom panel. Visibility errors across K band are between 3 and 5\%; H band data has been spectrally averaged to
	increase signal to noise ratio, the associated error is about 30\%. Solid lines correspond to the rim model (I), dashed lines to 
	the gas+rim model (II). Model parameters are reported in Tab.~1.
}
	\label{fig:viscl}
\end{figure*}

\begin{table}
\begin{center}
\begin{tabular}{l|l|l|l}
\hline
\hline
 & gas emission & dust emission &  \\
\hline
      &        & T$_{eff} \sim 1400$~K   &  incl = 40$^{\circ}$ \\ 	
      &	       & R$_{in}=0.34$~AU  &                \\
I     &        & R$_{out}=0.39$~AU &	            \\
      &        & X$_H=0.50$        &                \\
      &        & X$_K=0.79$        &                \\
\hline
      & T$_{eff}=2500$ K & T$_{eff}\sim1300$~K&  incl = 30$^{\circ}$  \\ 	
      &	R$_{in} \simgreat 0.05$~AU & R$_{in}=0.40$~AU  &            \\
II    & R$_{out} \simless 0.1$~AU & R$_{out}=0.49$~AU &	             \\
      & X$_H=0.41$       & X$_H=0.18$        &                 \\
      & X$_K=0.33$       & X$_K=0.46$        &                 \\
\hline
\hline
 \end{tabular}
\caption{Parameters of the best fit models shown in Fig.~\ref{fig:viscl} and \ref{fig:SED}.
	Model I corresponds to the rim model described in Isella and Natta (2005); in Model II we have added to the star and rim an additional source
of emission, as described in the text.
}
\label{tab:mod}
\end{center}
\end{table}

The results are presented in Fig.~\ref{fig:viscl}, which shows the visibilities and closure phases 
as function of wavelength for the two different data sets:
panel (a) refers to the Dec.30, 2006 observations (hour angle
 HA=0.25 h), panel (b) to the Jan.8, 2007 one (HA=0.54 h). 
The results from the two data sets  practically coincide, as expected given the 
small difference in HA. Due to the lower flux in the H band, data between 1.6 and 1.8 \um\ are averaged
in order to increase the signal to noise ratio. Even so, the error
on the H band visibility is about 30\%, much larger than that  
of the K band data ($\sim$5\%).

In the K band, the source is resolved on all the baselines, which 
provide an angular resolution of 1.8~mas, corresponding to 0.35~AU at
the distance of MWC~758. The closure phase is always larger than zero and increases 
with wavelength between 2 and 2.5 \um. Note that a non zero closure phase implies that 
the K band emission is not centrally symmetric. This measurement alone 
rules out models such as spherical envelopes, flat disks, etc. 
for the K band emission (see also the discussion in Monnier et al. 2006). 

In the H band, the MWC758 emission is unresolved on  the
UT3-UT4 and UT1-UT3 baselines, with a visibility equal to 1 within the error bars; 
on the longest baseline UT1-UT4 the visibility is slightly
smaller than unity for
both data sets (V$=0.8\pm0.1$). The H band closure phase is
consistent with 0, as expected given that the emission
is  practically unresolved.

\begin{figure*}[!t]
   \centering
   \subfigure[]
   {\includegraphics[width=9cm]{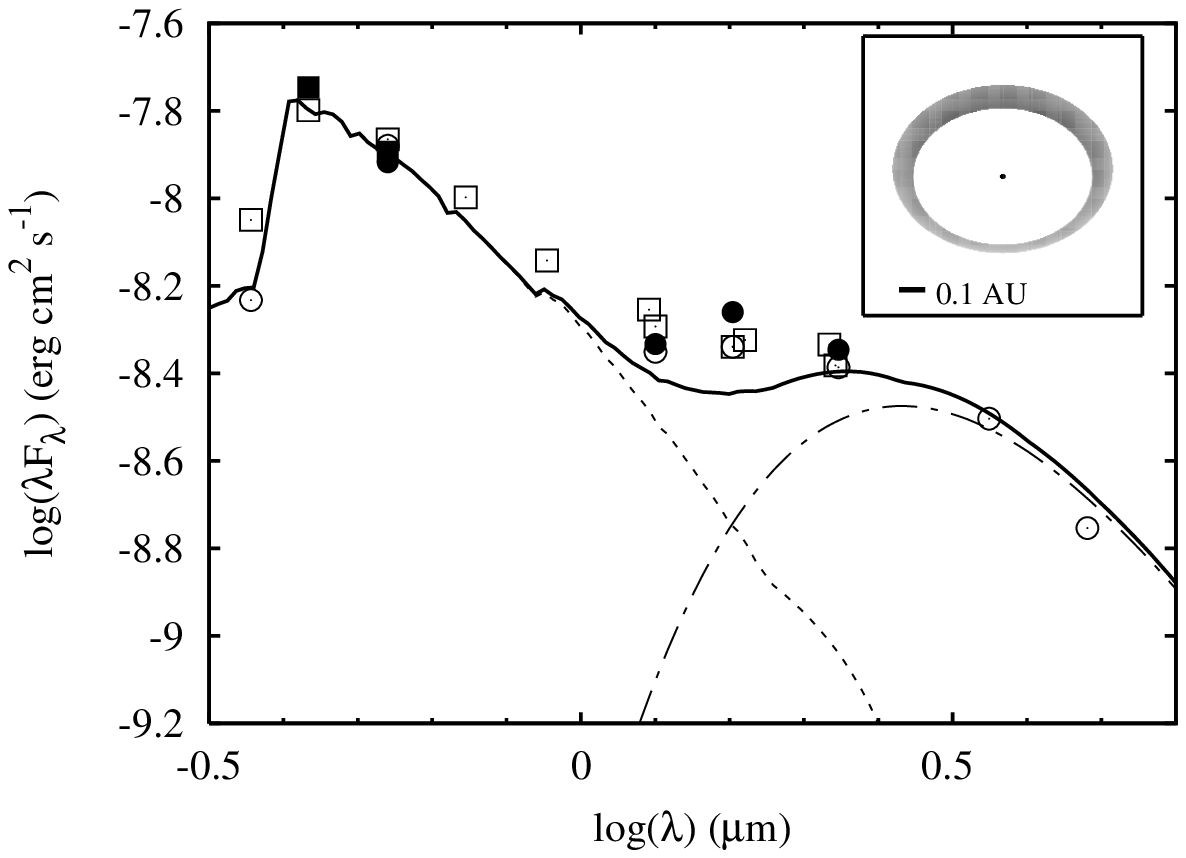}}
   \subfigure[]
   {\includegraphics[width=9cm]{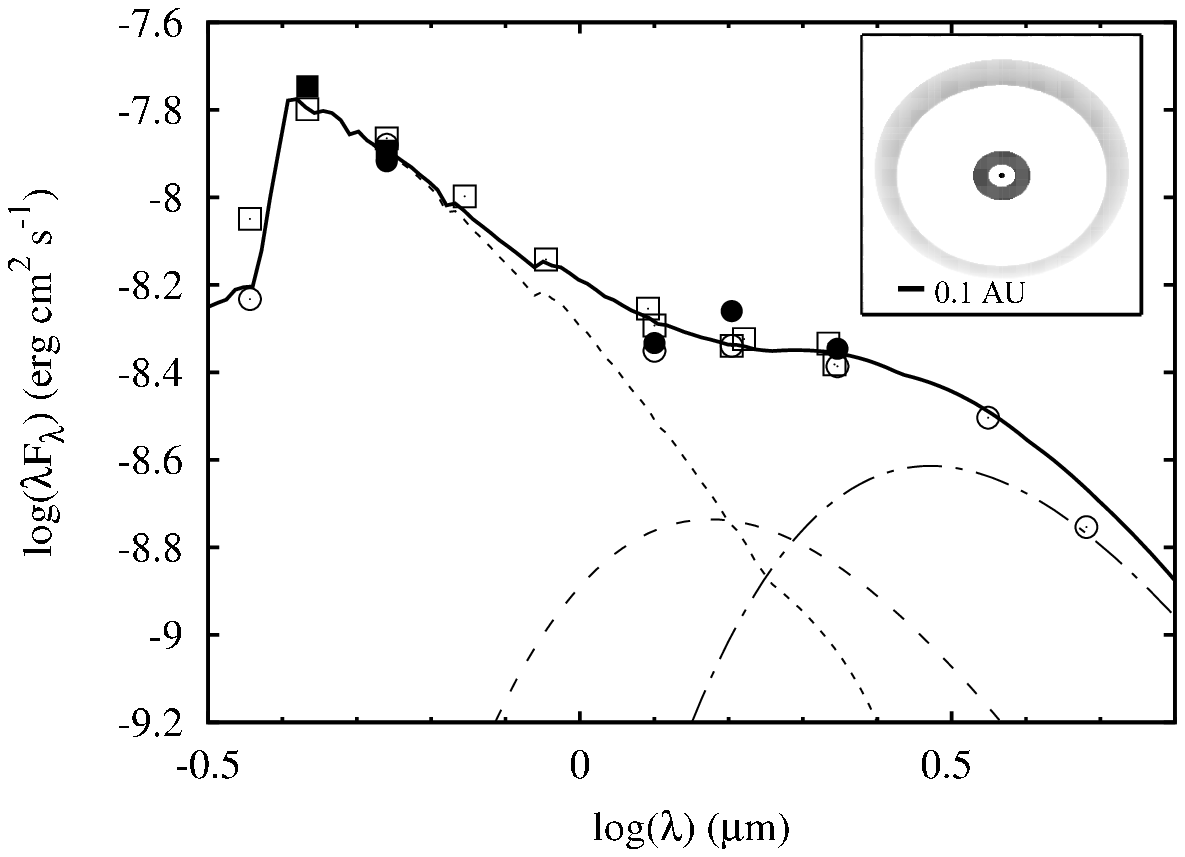}}
   \caption{SED of MWC~758. Symbols show the photometry from the literature (Eisner et al.~2004, Malfait et al.~1998, 
	Beskrovnaya et al.~1999, Cutri et al.~2003) corrected for an extinction $A_V=0.4$, R=3.1. The stellar spectrum 
	(short-dashed line) is that of a Kurucz model with \Teff=8000, Log $g$=4. Panel (a) shows the flux predicted by 
	model I (solid line) composed by the dusty disk rim emission (short-long dashed line) and the star. 
	Panel (b) shows the prediction for model II where an additional black body component is added (dashed line). 
	In each panel the image of the corresponding model is shown in the inset. Refer to Tab.~1 for model 
	parameters. }
\label {fig:SED}
\end{figure*}

\section {Discussion}

To analyze the AMBER data, we compute visibilities and closure phases as a
function of wavelength using the rim model of Isella and Natta (2005), where a complete description of the physical models and 
of their limitations can be found. Here we just want to remark that, once the stellar parameters are 
defined, the rim location and shape depend only on the grain emissivity, i.e., for known dust 
composition, on the grain size distribution. We assume silicate grains, which are the grains with the highest 
sublimation temperature according to Pollack et al.~(1994). The fitting procedure is described in 
Isella et al.~(2006) and the rim parameters that give the best fit to the data are given in Table 1.
R$_{in}$ is the distance of the rim from the star on the disk midplane, R$_{out}$
the distance of the top of the rim, so that $\Delta$R=R$_{out}-$R$_{in}$ 
measures the curvature of the rim due to the dependence of the dust sublimation 
temperature on the gas density. The rim emission can be approximately described
as a black-body at T$_{eff}$. The two quantities X$_H$ and X$_K$ give the fraction of
the dusty rim and the gaseous disk emission (see the discussion below) to the total flux in the H and K
band respectively; the missing flux to reach the unity is due to the stellar photosphere. 

\subsection {Star + rim models}

Model I includes the star and the rim emission only.
As shown in Fig.~\ref{fig:viscl}  (solid curves), it provides an adequate,
although not perfect, description of the 
K band visibility and closure phase measurements. Both the derived rim inner radius (0.34~AU) and
the disk inclination (40$^{\circ}$) are in agreement with the values obtained in Isella et al. (2006)
based on earlier PTI observations; the disk position angle is not well constrained by our data, varying between 
90$^{\circ}$ and 150$^{\circ}$. 

However, such models fail to account for the observations 
in the H band, where they predict that the total emission should
be spatially resolved.
The reason  is clear if we consider that 
the rim is physically narrow, e.g. 
$\Delta$R $\sim$0.05 AU, and  has practically the same size at all wavelengths. 
In this case, a variation of visibility and 
closure phase with wavelength may occur for two reasons only, one intrinsic, 
because the contribution of the unresolved stellar 
flux increases at shorter wavelengths, and one instrumental, since the interferometer resolution decreases with wavelength. 
In the calculation of the visibility these two effects tend to compensate one other and the resulting values 
are roughly constant, contrary to what we observe. The discrepancy between observations and model predictions is even more striking if we consider the dependence of
the closure phase on wavelength. The higher angular resolution achieved in H band amplifies 
the asymmetry of the model emission, resulting in a very large values
 of the closure phase in the H band, contrary to what is observed.
Note that a discrepancy between the rim model and the observations is also present in the SED (Fig.~\ref{fig:SED} panel (a)) 
since the best fit rim model can account only for about the 80\% of the emission measured in the H band.

\subsection {Evidence for gas emission in the inner disk}
\label{sec:gas}
Motivated by these considerations, we modified our rim model introducing an additional source 
of emission, which we describe as a uniform black body ring where the ring radii and its 
effective temperature are free parameters. The values of the parameters of the best-fitting model 
(Model II) are given in Table~1. As shown in Fig.~\ref{fig:viscl} and Fig.~\ref{fig:SED} 
(dashed lines) a good fit to both the H and K band visibilities and fluxes is obtained for 
T$_{eff}=2500$~K and ring outer radius $\simless 0.1$~AU. The exact values of the ring radii are 
not well constrained since the H band emission is poorly resolved. However, the lower H band 
visibility measured on the longest baseline seems to indicate ring radii between 0.05 and 0.1~AU. 
Fig.~2(b) shows that this model fits very well the MWC~758 SED. Such a model is clearly a simplified
representation of a more realistic inner disk in which the gas temperature and emission resonably 
decrease smoothly for increasing stellocentric distances. Nevertheless, it can reproduce the rough 
spatial scale for the gas emission and it is therefore adequate to understand the current observations.
 
\begin{figure}
   \centering
   \includegraphics[width=9cm]{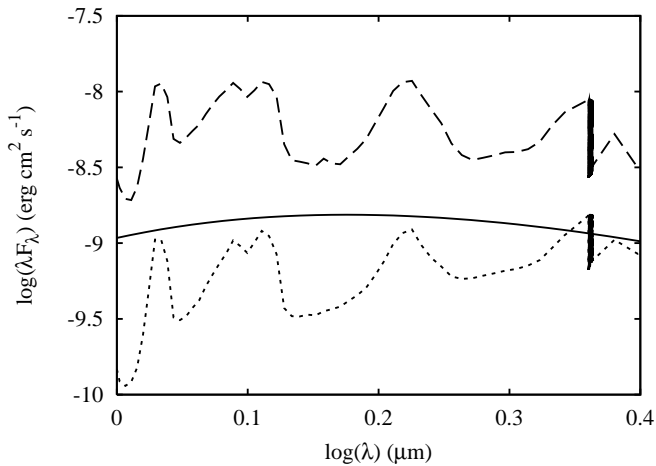}
   \caption{Comparison of the 2500 K black body emission, added in model II (solid line), with the
	emission of the inner gaseous disk  predicted by Muzerolle et al. (2004) 
  	for two mass accretion rates, $10^{-6}$ \Msun/yr (top curve) and $10^{-7}$ \Msun/yr 
	(lower curve), over the relevant reange of wavelength between 1 and 2.5 \um respectively.}
\label {fig:gas}
\end{figure}

The physical origin of this additional emission is hard to constrain from the available data. 
We suggest that it can be identified with the emission of the gaseous accretion disk inside 
the dusty rim, based on the comparison of the observational constraints we have with the predictions 
of the Muzerolle et al.~(2004) models. These authors model the physical structure of the inner 
disk of a typical Herbig Ae star, including  dust evaporation and the formation of the rim. 
Inside the rim, the gaseous disk properties depend on the mass accretion rate \Macc. 
We estimate that in MWC~758 \Macc $\sim 2\times 10^{-7}$ \Msun/yr, based on the strength of 
the \Brg\ emission\footnote{Telescopio Nazionale Galileo unpublished data}, as described in 
Garcia Lopez et al.~(2006). For such an \Macc, the models predict that the gas in the disk mid 
plane reaches a temperature of about 2500~K at a distance of ~0.1 AU, consistent with our constraint
that the hot component is located within 0.1~AU 
from the star. Fig.~\ref{fig:gas} shows the emission of the gas disk for two accretion rates
($10^{-7}$ and $10^{-6}$ \Msun/yr), which encompass the estimated value in MWC~758.
The 2500 K black body emission is roughly intermediate between the two, and its spectral shape is 
similar, supporting the interpretation of the ``second ring" as gas emission within the 
dust sublimation radius.

One should note that the Muzerolle et al.~(2004) models have been computed for slightly different 
stellar parameters, 
so that the comparison is necessarily very qualitative. 
Also, we were not able to compute the expected visibilities and phase closures
of the Muzerolle et al.~(2004) models, since only the wavelength dependence of the spatially integrated flux is available.

We finally remark that the introduction of the additional black body emission has
consequences for
the rim structure and for the derived dust grain size. In particular, going from model I to model II, 
the inner rim radius increases from 0.34~AU, obtained with a dust grain size of 1 \um, to 0.40~AU corresponding 
to dust grains of 0.5 \um. This confirms (Isella et al.~2006) that the grains
in the inner disk of MWC~758 are on average larger than the 
typical dust size, i.e. 0.01 \um, of grains in the interstellar medium. 
However,  an accurate determination of the grain properties
requires to understand the global properties of the inner disk, gas and dust.

\section {Conclusions}

In this Letter we present spectro-interferometric observations (R=35) of the pre-main 
sequence HAe star MWC~758, obtained with VLTI/Amber in the H and K bands. The observations 
have been performed with three UT telescopes in order to measure both 
spectrally dispersed  vibilities 
and closure phases. The H-band data and the closure phase are essential
quantities in order to distinguish between different models of
the disk structure. The observations provide a spatial 
resolution of 0.8 mas in the H band and 1.8 mas in the K band, corresponding to scale sizes 
of 0.16 and 0.35 AU, at the distance of MWC~758.  

In the K band, the source is resolved with non-zero closure phase (about 10$^{\circ}$),
as expected if the emission is due to dust in a rim at the sublimation radius (0.49~AU),
seen at an inclination of about 30$^{\circ}$. In the H band the source 
is only barely resolved with the longest baseline (125 m), implying that the emission at
these wavelengths arises within 0.1 AU from the central star.

We suggest that moving in wavelength from $\sim$2.5 $\mu$m to the H band, centered at
1.7 $\mu$m,  one moves from rim-dominated emission to a hotter, more compact component which 
is qualitatively consistent with the emission of the dust-free inner gaseous disk, as modeled 
by Muzerolle et al.~(2004). These results support the suggestion of Eisner et al.~(2007) 
that an extra component in addition to the dust would improve the fit to the spectrally 
dispersed, low resolution K band visibilities obtained with the Palomar Testbed Interferometer 
for a small sample of HAe stars. 

Our results show that the innermost gaseous disk can be detected and studied with current 
NIR interferometers through the detection of the continuum gas emission in the H band, where 
the gas contribution is higher than the stellar and dust emission, which
dominate at shorter and longer wavelengths, respectively. 
Acccurate models of the spatial and spectral dependence of the inner gaseous
disk  emission for a range of stellar and accretion properties, are
much needed.

\begin{acknowledgements}
We thank Paola d'Alessio and James Muzerolle for providing as with the gas SED models shown in Fig.~\ref{fig:gas}.
This project was partially supported by the INAF 2005 grant  ``Interferometria infrarossa: 
ottimizzazione di osservazioni astrofisiche" and by the INAF 2006 grant ``From Disks to Planetary Systems".
 This work was performed in part under contract with the Jet 
Propulsion Laboratory (JPL) funded by NASA through the Michelson Fellowship Program. 
JPL is managed for NASA by the California Institute of Technology

\end{acknowledgements}

{}



\begin{thebibliography}{}

\bibitem[2005]{Ak05}
Akeson, R. L.; Boden, A. F.; Monnier, J. D.; Millan-Gabet, R.; Beichman, C. et al., 2005, ApJ, 635,1173

\bibitem[1999]{B99}
Beskrovnaya, N. G.; Pogodin, M. A.; Miroshnichenko, A. S.; Th\'e, P. S.; Savanov, I. S.; Shakhovskoy, N. M.; Rostopchina, A. N.; Kozlova, O. V.; Kuratov, K. S.; 1999, A\&A, 343, 163

\bibitem[2007]{Br07} 
Brittain, Sean D.; Simon, Theodore; Najita, Joan R.; Rettig, Terrence W., 2007, ApJ, 659,685

\bibitem[2003]{C03}
Cutri, R. M.; Skrutskie, M. F.; van Dyk, S.; Beichman, C. A.; Carpenter, J. M.; Chester, T.; Cambresy, L.; Evans, T. et al.; 2003, The IRSA 2MASS All-Sky Point Source Catalog, NASA/IPAC Infrared Science Archive.

\bibitem[2001]{DDN01} 
Dullemond, C.P.; Dominik, C.; Natta, A.; 2001, ApJ 560, 957 

\bibitem[2004]{E04} 
Eisner, J.A.; Lane, B.F.; Hillebrand, L.A.; Akeson, R.L.; Sargent,
A.I.; 2004, ApJ 613, 1049 

\bibitem[2007]{E07}
Eisner, J. A.; Chiang, E. I.; Lane, B. F.; Akeson, R. L.; 2007, ApJ, 657, 347

\bibitem[2007]{E07b}
Eisner, J. A., 2007, Nature, 447, 7144

\bibitem[2006]{GL06}
Garcia Lopez, R.; Natta, A.; Testi, L.; Habart, E.; 2006, A\&A, 459, 837

\bibitem[2005]{I05}
Isella, A.; Natta, A.; 2005, A\&A, 438, 899I

\bibitem[2006]{I06}
Isella, A.; Testi, L.; Natta, A.; 2006, A\&A, 451, 951

\bibitem[2007]{M07}	
Malbet, F.; Benisty, M.; de Wit, W.-J.; Kraus, S.; Meilland, A. et al.; 2007, A\&A,
464, 43 

\bibitem[1998]{M98}
Malfait, K.; Bogaert, E.; Waelkens, C.; 1998, A\&A, 331, 211

\bibitem[2007]{Mi07}
Millan-Gabet, R.; Malbet, F.; Akeson, R.; Leinert, C. et al.; 2007,
Protostars and Planets V, B. Reipurth, D. Jewitt, and K. Keil (eds.), University of 
Arizona Press, Tucson, 951 pp., p.539-554

\bibitem[2006]{Mo06}
Monnier, J. D.; Berger, J.-P.; Millan-Gabet, R. et al, 2006, ApJ, 647, 444

\bibitem[2004]{MAC04} 
Muzerolle J., D'Alessio P., Calvet N., Hartmann L. 2004, ApJ, 617, 406 

\bibitem[2007]{Na07}
Najita, J. R.; Carr, J. S.; Glassgold, A. E.; Valenti, J. A., 2007, Protostars and Planets 
V, B. Reipurth, D. Jewitt, and K. Keil (eds.), University of Arizona Press, Tucson, 951 pp., p.507-522

\bibitem[Pasinetti Fracassini et al.(2001)]{cadar_1} 
Pasinetti Fracassini, L.~E., Pastori, L., Covino, S., \& Pozzi, A.\ 2001, \aap, 367, 521

\bibitem[1994]{PH94}
Pollack J.B., Hollenbach D., Beckwith S., Simonelli D.P., Roush T.,
Fong W. 1994, ApJ 421, 615 

\bibitem[Tatulli et al.(2007)]{tatulli_1} 

Tatulli, E.; Millour, F.; Chelli, A.; Duvert, G.; Acke, B et al.; 2007, \aap, 464, 29




\end{thebibliography}
\end{document}